\documentstyle[preprint,aps]{revtex}
\begin{document}
\draft
\preprint{SNUTP--94--62
\hspace{-38.0mm}
\raisebox{2.4ex}{KAIST--CHEP--94/04}}
\title{ $K_{L e5}$ decay as a background \\ in search for
$ K_{L} \rightarrow \pi^{0} \mu^{\pm} e^{\mp}$ }
\author{
Pyungwon Ko$^{a}$\thanks{pko@phya.snu.ac.kr},
Kang Young Lee$^{b}$\thanks{kylee@chep5.kaist.ac.kr}
and Jae Kwan Kim$^{b}$
}
\vspace{.5in}
\address{
$^{a}$ Dept. of Physics, Hong-Ik University,  Seoul 121-791,
KOREA  \\
$^{b}$ Dept. of Physics, KAIST, Taejon 305-701, KOREA
}
\maketitle
\begin{abstract}
We consider a process $K_{L e5} ( K_{L} \rightarrow \pi^{0} \pi^{0}
\pi^{\pm} e^{\mp} \nu )$ as a standard model  background to the experiment
$K_{L} \rightarrow \pi^{0} \mu^{\pm} e^{\mp}$, which seeks for possible
violation of  lepton family number.  Using the lowest order chiral
lagrangian, we find that the branching ratio for
$K_{L} \rightarrow \pi^{0} \pi^{0} \pi^{\pm} e^{\mp} \nu $  to be $6.2
\times 10^{-12}$.  A similar decay $K_{L} \rightarrow \pi^{\mp} \pi^{\mp}
\pi^{\pm} e^{\pm} \nu$ has a branching ratio, $1.7 \times
10^{-11}$.
\\
\end{abstract}
\pacs{ }

\narrowtext
\tighten


In the standard model, neutrinos are assumed to be massless so that the
individual lepton numbers  are exactly conserved separately.  Thus,
such decays as $K_{L} \rightarrow \mu^{\pm} e^{\mp}$ and $K \rightarrow
\pi \mu^{\pm} e^{\mp}$ are not allowed in the standard model (SM).
However, this is no longer true in many models beyond SM where neutrinos
are endowed with small masses.   These small neutrino masses can generate
lepton flavor oscillations, lepton family number violations, and may
explain the solar neutrino problem and the atmospheric problem
simultaneously.

In view of these, it is important to look for lepton
family number violating processes  such as $K \rightarrow \pi
\mu^{\pm} e^{\mp}$ and $K_{L} \rightarrow \mu^{\pm} e^{\mp}$, independently
of other neutrino experiments.    Up to now, only $K^{\pm} \rightarrow
\pi^{\pm} \mu^{\pm} e^{\mp} $ and $K_{L} \rightarrow \mu^{\pm} e^{\mp} $
have been searched for :
\begin{eqnarray}
B ( K^{\pm} \rightarrow \pi^{\pm} \mu^{\pm} e^{\mp}) & < & 2.1
\times 10^{-10}~~~({\rm AGS \cite{bags}}),
\\
B ( K_{L} \rightarrow \mu^{\pm} e^{\mp} ) & < & 9.7 \times 10^{-11}~~~
({\rm KEK-137 \cite{kek}}),
\\
& & 3.3 \times 10^{-11} ~~~({\rm AGS-791 \cite{ags}}).
\end{eqnarray}
Another decay $K_{L} \rightarrow \pi^{0} \mu^{\pm} e^{\mp} $ is being
studied at FNAL for the first time.  This decay is related with
$ K^{\pm} \rightarrow \pi^{\pm} \mu^{\pm} e^{\mp}$ via isospin rotation,
if the quark family number is unbroken, implying that
$B (K_{L} \rightarrow \pi^{0} \mu^{\pm} e^{\mp}) < 4 \times 10^{-10}$.
However, the quark family number may be broken in beyond standard models
just as the lepton family number could be broken.  Therefore,  a
search for  $K_{L} \rightarrow \pi^{0} \mu^{\pm} e^{\mp}$ can provide
independent informations on the lepton family number violation as a
possible signal due to new physics.

Typically, the branching ratio for   $K_{L} \rightarrow \mu^{\pm} e^{\mp}$
in beyond SM is an order of $\sim O(10^{-18})$ \cite{marciano}.
Thus, the branching ratio for $K_{L} \rightarrow \pi^{0}
\mu^{\pm} e^{\mp}$ should be an order of $\sim O(10^{-19 \sim -20})$,
because {\it (i)}
the phase  space is less  and {\it (ii)} the Nambu--Goldstone nature of
a pion produces  one more  power of the pion momentum in the amplitude
compared to  $K_{L} \rightarrow \pi^{0} \mu^{\mp} e^{\pm}$.

The expected branching ratio for $K_{L} \rightarrow \pi^{0} \mu^{\pm}
e^{\mp}$ being so small, one can expect that there would be several
sources of backgrounds from SM which can mimic this decay.
For example, semileptonic $ K_{e4}$ and $K_{e5}$ decays may become
backgrounds when a charged pion is misidentified as a muon.
However, these possibilities can be easily overcome by imposing a constraint
that the invariant mass of the $\pi^{\pm} e^{\mp}$ and $\pi^{0}$ (two
photons with $m_{\gamma\gamma} = m_{\pi^0}$ in actualty) is around
$m_{K}$ with an experimental uncertainty, typically $m_{K} \pm 10$ MeV.
However, there is an exception for this argument in the case of $K_{L e5}$.
In $K_{L} \rightarrow \pi^{0} \pi^{0} \pi^{\mp} e^{\pm} \nu$, there arise
four photons decaying from two $\pi^0$'s, and two photons (each from
different parental $\pi^0$) could escape detection and remaining two
photons may have the invariant mass around $m_{\pi^0}$ accidentally.
To estimate this kind of background from $K_{L e5}$, we need to know
the amplitude for this process as well as experimental settings.
The first decay $K_{e4}$ have been studied in detail  both
theoretically and experminetally, whereas  the $K_{L e5}$ decay
has not been discussed in the literature.

In this paper, we study a process $K_{e5}$ as a possible standard model
background  to an experiment searching for $K_{L} \rightarrow \pi^{0}
\mu^{\pm} e^{\mp}$ using the lowest order chiral perturbation theory.
Our amplitude can be used for a background estimate in search for
$K_{L} \rightarrow \pi^{0} \mu^{\mp} e^{\pm}$ as described in the
previous paragraph.  For completeness, we also include another decay
$K_{L} \rightarrow \pi^{\mp} \pi^{\mp} \pi^{\pm} e^{\pm} \nu$ at the end.

Strong and electromagnetic interactions among pion and kaons and their weak
decays are well described in terms of the chiral
lagrangian invariant under  $SU(3)_{L} \times SU (3)_{R}$
transformations as long as their four--momenta are not too large.
Since $m_{K_L} = 498$ MeV and $m_{\pi} =  140$ MeV,
each pion momentum is very small in the $K_{e5}$ decay.
Hence it is sufficient to consider the lowest order lagrangian only.
At the lowest order $O(p^2)$, the $SU(3)_{L} \times SU(3)_{R}$
chiral lagrangian is  ~\cite{gasser}
\begin{equation}
{\cal L} = \frac{f_{\pi}^2}{4}~
	   \mbox{Tr} ~(D_{\mu}\Sigma D^{\mu}\Sigma^{\dagger})
	 + \frac{f_{\pi}^2}{2}~
	   \mbox{Tr} ~(\mu M \Sigma + \Sigma^{\dagger} \mu M)
\end{equation}
where
\begin{eqnarray}
D_{\mu} \Sigma &=& \partial_{\mu} \Sigma
		     -i r_{\mu} \Sigma + i \Sigma ~l_{\mu},     \\
	\Sigma &=& \exp(\frac{2i}{f_{\pi}} \phi),
\end{eqnarray}
$f_{\pi} = 93 $ MeV is the pion decay constant,
and $\phi$  is the pseudoscalar Nambu--Goldstone boson matrix,
\begin{eqnarray*}
\phi = {1\over \sqrt{2}}~\left( \begin{array}{ccc}
	   \pi^{0} / \sqrt{2} + \eta / \sqrt{6} &
	    \pi^{+} & K^{+}  \\
	    \pi^{-} & -\pi^{0} / \sqrt{2} + \eta / \sqrt{6} &
	    K^{0} \\
	    K^{-} & \bar{K^0} & - 2 \eta / \sqrt{6}
	   \end{array} \right),
\end{eqnarray*}
and $M = {\rm diag} (m_{u}, m_{d}, m_{s})$ is the current quark mass matrix.
The matrix field $\Sigma$ transforms as $\Sigma \rightarrow R \Sigma
L^{\dagger}$ under chiral $SU(3)_{L} \times SU(3)_R$ transformations.

The external gauge fields $l_{\mu}, r_{\mu}$ are appropriate linear
combinations  of $SU (2)_{L} \times U (1)_{Y}$ electroweak gauge fields
of the standard model.
Since we are not interested in electromagetic interactions in this work,
we may set
\begin{eqnarray}
l_{\mu} & = & -\frac{g}{\sqrt{2}} W_{\mu}^- T,
\nonumber \\
r_{\mu} & = & 0,
\end{eqnarray}
where
\begin{eqnarray*}
T = \left( \begin{array}{c}
	   ~0~0~V_{us}\\
	    0~0~0~     \\
	    0~0~0~
	   \end{array} \right).
\end{eqnarray*}

The amplitude for $K^{0} \rightarrow (3 \pi)^{-} e^+ \nu$  can be written as
\begin{equation}
{\cal M} (K^{0} \rightarrow  (3 \pi)^{-} l^{+} \nu) \equiv
{G_{F} \over \sqrt{2}}~V_{us}~\langle (3 \pi)^{-} | \bar{s} \gamma_{\mu}
( 1 - \gamma_{5} ) u | K^{0} \rangle~\bar{u}_{\nu} \gamma^{\mu} ( 1 -
\gamma_{5} ) v_{l}.
\end{equation}
The matrix element of the $(V-A)$ hadronic current
\begin{equation}
H_{\mu}^{abc} (p_{1},p_{2},p_{3}) \equiv  \langle \pi^{a} (p_{1}) \pi^{b}
(p_{2}) \pi^{c} (p_{3})  | \bar{s} \gamma_{\mu} ( 1 - \gamma_{5} ) u
| K^{0} (k) \rangle
\label{eq:hmu}
\end{equation}
can be read off from the gauged chiral lagrangian, Eq.~(4), as coefficients
of $g V_{us} W_{\mu}^{-} /  2 \sqrt{2}$.
In Eq.~(\ref{eq:hmu}), the superscripts  $a,b,c$ denote the electric charge
of each pion, and the third pion is the charge odd one as usual.
The square of the matrix element summed over the spins of the final leptons
is given by (suppressing the superscripts of $H_{\mu}$ for the moment)
\begin{equation}
| \bar{\cal M} |^{2} = {G_{F}^2 \over 2}~| V_{us} |^{2} ~H_{\rho\sigma}~
L^{\rho\sigma},
\end{equation}
where
\begin{eqnarray}
H_{\rho\sigma} & = & H_{\rho} H_{\sigma}^{*},
\\
L_{\rho\sigma} & = &  8 \left[ p_{e \rho} p_{\nu \sigma} + p_{\nu \rho} p_{e
\sigma} - g_{\rho \sigma} p_{e} \cdot p_{\nu} + i \epsilon_{\rho \sigma
\alpha \beta}~p_{e}^{\alpha} p_{\nu}^{\beta} \right].
\end{eqnarray}

Let us first consider $K^{0} \rightarrow \pi^{0} \pi^{0} \pi^{-} e^{+} \nu$.
It is straightforward to extract from Eq.~(4)
the following interaction lagrangian relevant to the $K^{0} \rightarrow
\pi^{0} \pi^{0} \pi^{-} e^{+} \nu$ decay :
\begin{eqnarray}
{\cal L}_{s} & = &
	     -\frac{1}{12 f_{\pi}^2}
	      \left( \pi^0 \pi^0 \partial_{\mu}K^0 \partial^{\mu}{\bar K}^0
	      + {\bar K}^0 K^0 \partial_{\mu}\pi^0 \partial^{\mu}\pi^0
	      - \pi^0 K^0 \partial_{\mu}\pi^0 \partial^{\mu}{\bar K}^0
	      - \pi^0 {\bar K^0} \partial_{\mu}\pi^0 \partial^{\mu}K^0
	      \right)
	      \nonumber \\
	   && + \frac{\sqrt{2}}{4 f_{\pi}^2}
		\left( \pi^0 K^0 \partial_{\mu}K^- \partial^{\mu}\pi^+
	      + \pi^+ K^- \partial_{\mu}\pi^0 \partial^{\mu}K^0
	      - \pi^0 K^- \partial_{\mu}\pi^+ \partial^{\mu}K^0
	      - \pi^+ K^0 \partial_{\mu}\pi^0 \partial^{\mu}K^-
	      \right)
	      \nonumber \\
           && + { 1\over 3 f_{\pi}^2}
                \left( \pi^{0} \pi^{+} \partial_{\mu} \pi^{-} \partial^{\mu}
                 \pi^{0} + \pi^{0} \pi^{-} \partial_{\mu} \pi^{+}
                 \partial^{\mu} \pi^{0} - \pi^{0} \pi^{0} \partial_{\mu}
                  \pi^{+} \partial^{\mu} \pi^{-} - \pi^{+} \pi^{-}
                 \partial_{\mu} \pi^{0} \partial^{\mu} \pi^{0} \right)
               \\
	   && + \frac{1}{12 f_{\pi}^2}
		\left(\mu (3m_d+m_s) K^0{\bar K}^0 \pi^0 \pi^0
		  + \sqrt{2} \mu (m_u-m_d) K^0 K^- \pi^0 \pi^+
               + 2 \mu (m_u+m_d) \pi^0 \pi^0 \pi^+ \pi^- \right)
	      \nonumber     \\
{\cal L}_{w} & = &
	       -i\frac{g}{2\sqrt{2}} V_{us} {W^{-}}^{\mu}
	       \left[
		\left( \partial_{\mu} \pi^{+} K^0
			 - \partial_{\mu} K^0 \pi^{+}  \right)
	   + \frac{1}{\sqrt{2}}
	       \left( \partial_{\mu} \pi^{0} K^+
			 - \partial_{\mu} K^+ \pi^{0} \right)
	       \right.
	   \nonumber \\
	   && ~~~~~~~~~~~~~~~~~~~\left. -\frac{1}{12 f_{\pi}^2}
	       \left( 7\partial_{\mu} \pi^{+} \pi^0 \pi^0 K^0
		 -6\partial_{\mu} \pi^{0} \pi^0 \pi^+ K^0
		 - \partial_{\mu} K^0 \pi^0 \pi^0 \pi^{+} \right)
	      \right]
\end{eqnarray}
In this work, we ignore isospin symmetry breaking due to  $m_{u} \neq m_{d}$.
Therefore, $ \mu ( 3 m_{d} + m_{s} ) \simeq (m_{K}^{2}+ m_{\pi}^{2})$ with
$m_{K} = 498$ MeV, and $m_{\pi} = 135$ MeV, {\it etc.}.

Feynman diagrams relevant to  the $K^{0} \rightarrow \pi^{0}
\pi^{0} \pi^{-}  e^{+} \nu $ decay derived
from the above  lagrangians are shown in Fig.~1.
The closed circle and the square blob denote  strong and weak vertices,
respectively.
Evaluating the Feynman diagrams shown in Fig.~1,   one gets
\begin{equation}
H_{\mu}^{00-} (p_{1},p_{2},p_{3}) = \Sigma_{i=a}^{d} {\cal M}_{\mu}^{(i)},
\end{equation}
where
\begin{eqnarray}
{\cal M}_{\mu}^{(a)} & = & {1 \over 6 f_{\pi}^2}~
\left[ 2 ( p_{1} + p_{2} )_{\mu} - 8 p_{3 \mu} + L_{\mu} \right],
\\
{\cal M}_{\mu}^{(b)} & = & - {1 \over 3 f_{\pi}^2}~
{ ( 2 q_{\mu} + L_{\mu} ) \over q^{2} - m_{\pi}^2 }~\left[ m_{\pi}^{2}
+ 2 ( 2 p_{1} \cdot p_{2} - p_{1} \cdot p_{3} - p_{2} \cdot p_{3} ) \right],
\\
{\cal M}_{\mu}^{(c)} & = & -{1 \over 6 f_{\pi}^2}~{
( 2 p_{3\mu} + L_{\mu} ) \over (k - p_{1} - p_{2} )^{2} - m_{K}^{2} }
{}~\left[
k \cdot (p_{1} + p_{2}) + (p_{1} + p_{2})^{2}
\right],
\\
{\cal M}_{\mu}^{(d)} & = & - {1 \over 2 f_{\pi}^2}~{
( 2 p_{2\mu} + L_{\mu} ) \over ( k - p_{1} - p_{3} )^{2} - m_{K}^{2} }~
{}~ [ k \cdot (p_{1} - p_{3}) ] ~
+ ( p_{1} \leftrightarrow p_{2} ),
\end{eqnarray}
with $ L = ( p_{e} + p_{\nu} ),~~ q  = ( p_{1} + p_{2} + p_{3} ) =
(k - L)$.
Thus, to the lowest order in chiral expansion, the matrix element of the
$(V-A)$ hadronic current in $K^{0} \rightarrow \pi^{0} \pi^{0} \pi^{+}
e^- \nu$ depend on four independent form factors, $F_{i} ~(i = 1,2,3,4)$ :
\begin{equation}
H_{\mu}^{00-} =
F_{1} p_{1 \mu} + F_{2} p_{2 \mu} + F_{3} p_{3 \mu} + F_{4} L_{\mu} .
\label{eq:hmu1}
\end{equation}
In actualty, the last form factor $F_{4}$ can be ignored in the following
since its contribution is proportional to the lepton mass ($m_{e}$ in  the
present case).


In order to get the physical quantities from the amplitude obtained above,
we have to perform the 5--body phase space integration.
First of all, let us write the differential decay rate for $K_{L e5}$ decay
as
\begin{equation}
d \Gamma (K_{L} \rightarrow  \pi^{0} \pi^{0} \pi^{\pm} e^{\mp} \nu)
= {1 \over 2 m_K (2 \pi)^{11}}~{G_{F}^{2} \over 2}~| V_{us}|^{2}~
H_{\rho\sigma}^{00 \pm} L^{\rho\sigma}
d_{5} (PS),
\label{eq:rate}
\end{equation}
Here, the decay rate includes sum over all possible charge states of the
final lepton.
In order to take into account the two identical $\pi^0$'s in the final
state, we have to  divide the final result obtained from (\ref{eq:rate})
by factor of 2.
The five body phase space $d_{5} (PS)$  is defined as
\begin{equation}
d_{5} (PS) = \delta^{4} (k - \Sigma_{i} p_{i}) ~\Pi_{i} {d^{3} \vec{p}_{i}
\over 2 E_{i}},
\end{equation}
and can be expressed in terms of products of reduced two body phase space
in a standard manner.
The five body phase space integration in Eq.~(\ref{eq:rate})
was numerically performed, and we get
\begin{equation}
B(K_{L} \rightarrow \pi^0 \pi^0 \pi^{\pm} e^{\mp} \nu) =  6.2 \times 10^{-12}.
\end{equation}
Being this small, the possibility that $K_{Le5}$ might be a background to
$K_{L} \rightarrow \pi^{0} \mu^{\pm} e^{\mp}$ may be negligible at the
current stage of experiments.  However, one needs more detailed simulations
of the actual experiment using our amplitude Eqs.~(16)--(19) to be more
definite.

For completeness, we consider $K_{L} \rightarrow \pi^{+} \pi^{-} \pi^{\pm}
e^{\mp} \nu$.
There are only three Feynman diagrams which contribute to this decay
as shown in Figs.~1 (a)--(c) in paretheses.
The interaction lagrangian relevant to this decay is
\begin{equation}
{\cal L}_{w} =  i { g V_{us} \over 12 \sqrt{2} f_{\pi}^2}~W^{- \mu}~
	      \left[ 4 \partial_{\mu} \pi^{+} K^{0} \pi^{+} \pi^{-}
		 - 3 \partial_{\mu} \pi^{-} K^{0} \pi^{+} \pi^{+}
		 - \partial_{\mu} K^{0} \pi^{+} \pi^{+} \pi^{-}
	      \right],
\label{eq:lweak2}
\end{equation}
as well as Eq.~(14), and
\begin{eqnarray}
{\cal L}_{s} & = & { 1 \over 6 f_{\pi}^2}~\left[ \pi^{-} \pi^{-}
\partial_{\mu} \pi^{+} \partial^{\mu} \pi^{+}
+ \pi^{+} \pi^{+} \partial_{\mu} \pi^{-} \partial^{\mu} \pi^{-}
- 2 \pi^{+} \pi^{-} \partial_{\mu} \pi^{+} \partial^{\mu} \pi^{-}
+ \mu ( m_{u} + m_{d} )
\pi^{+} \pi^{+} \pi^{-} \pi^{-} \right]
\nonumber
\\
& & + {1 \over 6 f_{\pi}^{2} }~\left[ 2 \left( \pi^{+} \bar{K}^{0}
\partial_{\mu} \pi^{-} \partial^{\mu} K^{0} + \pi^{-} K^{0}
\partial_{\mu} \pi^{+} \partial^{\mu} \bar{K}^{0} \right) - \pi^{-}
\bar{K}^{0} \partial_{\mu} \pi^{+} \partial^{\mu} K^{0} - \pi^{+} K^{0}
\partial_{\mu} \pi^{-} \bar{K}^{0} \right.
\nonumber   \\
&& \left. ~~~~~~~~ - \pi^{+} \pi^{-} \partial_{\mu} K^{0}
\partial^{\mu} \bar{K}^{0} - K^{0} \bar{K}^{0} \partial_{\mu} \pi^{+}
\partial^{\mu} \pi^{-} + \mu (m_{u} + 2 m_{d} + m_{s} ) K^{0} \bar{K}^{0}
\pi^{+} \pi^{-}
\right]
\label{eq:lstrong2}
\end{eqnarray}
Calculating these diagrams with the interaction lagrangians  given in
Eqs.~(9),(\ref{eq:lweak2}) and (\ref{eq:lstrong2}), we get
\begin{equation}
H_{\mu}^{--+} (p_{1},p_{2},p_{3}) = \Sigma_{i=a}^{c} {\cal M}_{\mu}^{(i)},
\end{equation}
where
\begin{eqnarray}
{\cal M}_{\mu}^{(a)} & = & -{1 \over 3 f_{\pi}^2}~
\left[ 3 ( p_{1} + p_{2} )_{\mu} - 2 p_{3 \mu} + L_{\mu} \right],\\
{\cal M}_{\mu}^{(b)} & = & - {2 \over 3 f_{\pi}^2}~{
( 2 q_{\mu} + L_{\mu} ) \over q^{2} - m_{\pi}^2 }~\left[
m_{\pi}^{2} - 2 p_{1} \cdot p_{2} + p_{1} \cdot p_{2} + p_{2} \cdot p_{3}
 \right],
\\
{\cal M}_{\mu}^{(c)} & = & -{1 \over 3 f_{\pi}^2}~
{( 2 p_{1 \mu} + L_{\mu} ) \over (p_{1} + L)^{2} - m_{K}^{2}}~\left[
m_{\pi}^{2} + p_{1} \cdot p_{3} + k \cdot (2 p_{3} - p_{1} )
\right]
\\
&& + ( p_{1} \leftrightarrow p_{2} ).  \nonumber
\end{eqnarray}
Summing over the all possible charges of the final lepton and performing
the phase space integrations numerically, we get
\begin{equation}
B ( K_{L} \rightarrow \pi^{\pm} \pi^{\pm} \pi^{\mp} e^{\mp} \nu )
= 1.7 \times 10^{-11}.
\end{equation}


In conclusion,   the amplitudes for $K_{L e5}$ decays are  derived using
the lowest order chiral perturbation theory, which can be used by
experimentalists in order to study the background to $K_{L} \rightarrow
\pi^{0} \mu^{\pm} e^{\mp}$ coming from the decay, $K_{L} \rightarrow \pi^{0}
\pi^{0} \pi^{\pm} e^{\mp} \nu$.
The resulting branching ratio for a decay
$K_{L} \rightarrow \pi^{0} \pi^{0} \pi^{\pm} e^{\mp} \nu$ is about
$6.2 \times 10^{-12}$.  The branching ratio for a similar decay
$K_{L} \rightarrow \pi^{\mp} \pi^{\mp} \pi^{\pm} e^{\pm} \nu$ is twice
larger, $1.7 \times 10^{-11}$.
Thus, these decays are unlikely to be observed in the current and near-future
experiments.

\acknowledgements

P.K. thanks  Prof. J.L. Rosner for suggesting this problem and useful
correspondence. We are also grateful to Prof. S.K. Kim for discussions
about the experimental situations.
P.K. was supported in part by KOSEF  through CTP at Seoul  National
University, and by the Basic Science Research Institute Program, Ministry
of Education, 1994, Project No. BSRI--94--2425.
K.Y.Lee and J.K. Kim were supported by KOSEF.

%
%

\begin{figure}
\caption{
Feynman diagrams  relevant to $K^{0} \rightarrow \pi^{0} \pi^{0} \pi^{-}
e^{+} \nu $ in chiral perturbation theory to
$O(p^{2})$.  The square blob and the closed circle represent strong and
weak interaction vertices, respectively.
(The diagrams for $K^{0} \rightarrow \pi^{-} \pi^{-} \pi^{+} e^{+} \nu$ in
he parentheses in (a), (b) and (c). There is no diagram anologous to (d)
contributing to $K^{0} \rightarrow \pi^{-} \pi^{-} \pi^{+} e^{+} \nu$. )
}
\label{figone}
\end{figure}

\end{document}